
\catcode`@=11

\def\singlespace{\normalbaselines}
\def\oneandahalfspace{\baselineskip=1.15\normalbaselineskip plus 1pt
\lineskip=2pt\lineskiplimit=1pt}

\def\np{\vfill\eject}
\def\nl{\hfil\break}

\def\nofirstpagenoten{\nopagenumbers\footline={\ifnum\pageno>1\tenrm
\hss\folio\hss\fi}}
\def\nofirstpagenotwelve{\nopagenumbers\footline={\ifnum\pageno>1\twelverm
\hss\folio\hss\fi}}
\def\leaderfill{\leaders\hbox to 1em{\hss.\hss}\hfill}
\def\ft#1#2{{\textstyle{{#1}\over{#2}}}}
\def\frac#1#2{\leavevmode\kern.1em
\raise.5ex\hbox{\the\scriptfont0 #1}\kern-.1em/\kern-.15em
\lower.25ex\hbox{\the\scriptfont0 #2}}
\def\sfrac#1#2{\leavevmode\kern.1em
\raise.5ex\hbox{\the\scriptscriptfont0 #1}\kern-.1em/\kern-.15em
\lower.25ex\hbox{\the\scriptscriptfont0 #2}}

\parindent=20pt
\def\narrow{\advance\leftskip by 40pt \advance\rightskip by 40pt}

\def\AB{\bigskip
        \centerline{\bf ABSTRACT}\medskip\narrow}
\def\nonarrower{\advance\leftskip by -40pt\advance\rightskip by -40pt}
\def\AE{\bigskip\nonarrower}

\def\boxit#1{\vbox{\hrule\hbox{\vrule\kern3pt
        \vbox{\kern3pt#1\kern3pt}\kern3pt\vrule}\hrule}}

\def\gtorder{\mathrel{\raise.3ex\hbox{$>$}\mkern-14mu
             \lower0.6ex\hbox{$\sim$}}}
\def\ltorder{\mathrel{\raise.3ex\hbox{$<$}|mkern-14mu
             \lower0.6ex\hbox{\sim$}}}
\def\dalemb#1#2{{\vbox{\hrule height .#2pt
        \hbox{\vrule width.#2pt height#1pt \kern#1pt
                \vrule width.#2pt}
        \hrule height.#2pt}}}
\def\square{\mathord{\dalemb{4.9}{5}\hbox{\hskip1pt}}}

\font\twelvett=cmtt12 \font\twelvebf=cmbx12
\font\twelverm=cmr12  \font\twelvei=cmmi12
\font\twelvessr=cmss12 \font\twelvembi=cmmib10 scaled \magstep1
\font\twelvesy=cmsy10 scaled \magstep1
\font\twelvesl=cmsl12 \font\twelveex=cmex10 scaled \magstep1
\font\twelveit=cmti12
\font\tenssr=cmss10 \font\tenmbi=cmmib10
 
 \font\ninebf=cmbx9
\font\ninerm=cmr9  \font\ninei=cmmi9
\font\ninesy=cmsy9 \font\ninessr=cmss9
\font\ninembi=cmmib10 scaled 900
\font\eightit=cmti8 \font\eightsl=cmsl8
\font\eighttt=cmtt8 \font\eightbf=cmbx8
\font\eightrm=cmr8  \font\eighti=cmmi8
\font\eightsy=cmsy8 \font\eightex=cmex10 scaled 800
\font\eightssr=cmss8 \font\eightmbi=cmmib10 scaled 800
 
\font\sevenbf=cmbx7 \font\sevenrm=cmr7 \font\seveni=cmmi7
\font\sevensy=cmsy7 
\font\sevenssr=cmss9 scaled 778 \font\sevenmbi=cmmib10 scaled 700
\font\fivessr=cmss10 scaled 500  \font\fivembi=cmmib10 scaled 500

\newskip\ttglue
\newfam\ssrfam
\newfam\mbifam

\mathchardef\alpha="710B
\mathchardef\beta="710C
\mathchardef\gamma="710D
\mathchardef\delta="710E
\mathchardef\epsilon="710F
\mathchardef\zeta="7110
\mathchardef\eta="7111
\mathchardef\theta="7112
\mathchardef\iota="7113
\mathchardef\kappa="7114
\mathchardef\lambda="7115
\mathchardef\mu="7116
\mathchardef\nu="7117
\mathchardef\xi="7118
\mathchardef\pi="7119
\mathchardef\rho="711A
\mathchardef\sigma="711B
\mathchardef\tau="711C
\mathchardef\upsilon="711D
\mathchardef\phi="711E
\mathchardef\chi="711F
\mathchardef\psi="7120
\mathchardef\omega="7121
\mathchardef\varepsilon="7122
\mathchardef\vartheta="7123
\mathchardef\varpi="7124
\mathchardef\varrho="7125
\mathchardef\varsigma="7126
\mathchardef\varphi="7127
\mathchardef\partial="7140

\def\twelvepoint{\def\rm{\fam0\twelverm}
\textfont0=\twelverm \scriptfont0=\ninerm \scriptscriptfont0=\sevenrm
\textfont1=\twelvei \scriptfont1=\ninei \scriptscriptfont1=\seveni
\textfont2=\twelvesy \scriptfont2=\ninesy \scriptscriptfont2=\sevensy
\textfont3=\twelveex \scriptfont3=\twelveex \scriptscriptfont3=\twelveex
\def\it{\fam\itfam\twelveit} \textfont\itfam=\twelveit
\def\sl{\fam\slfam\twelvesl} \textfont\slfam=\twelvesl
\def\bf{\fam\bffam\twelvebf} \textfont\bffam=\twelvebf
\scriptfont\bffam=\ninebf \scriptscriptfont\bffam=\sevenbf
\def\tt{\fam\ttfam\twelvett} \textfont\ttfam=\twelvett
\def\ssr{\fam\ssrfam\twelvessr} \textfont\ssrfam=\twelvessr
\scriptfont\ssrfam=\ninessr \scriptscriptfont\ssrfam=\sevenssr
\def\mbi{\fam\mbifam\twelvembi} \textfont\mbifam=\twelvembi
\scriptfont\mbifam=\ninembi \scriptscriptfont\mbifam=\sevenmbi
\tt \ttglue=.5em plus .25em minus .15em
\normalbaselineskip=14pt
\bigskipamount=14pt plus4pt minus4pt
\medskipamount=7pt plus2pt minus2pt
\abovedisplayskip=14pt plus 3pt minus 10pt
\belowdisplayskip=14pt plus 3pt minus 10pt
\abovedisplayshortskip=0pt plus 3pt
\belowdisplayshortskip=8pt plus 3pt minus 5pt
\parskip=3pt plus 1.5pt
\tenfoot
\setbox\strutbox=\hbox{\vrule height10pt depth4pt width0pt}
\let\sc=\ninerm
\let\big=\twelvebig \normalbaselines\rm}
\def\twelvebig#1{{\hbox{$\left#1\vbox to10pt{}\right.\n@space$}}
\def\square{\mathord{\dalemb{5.9}{6}\hbox{\hskip1pt}}}}

\def\tenpoint{\def\rm{\fam0\tenrm}
\textfont0=\tenrm \scriptfont0=\sevenrm \scriptscriptfont0=\fiverm
\textfont1=\teni \scriptfont1=\seveni \scriptscriptfont1=\fivei
\textfont2=\tensy \scriptfont2=\sevensy \scriptscriptfont2=\fivesy
\textfont3=\tenex \scriptfont3=\tenex \scriptscriptfont3=\tenex
\def\it{\fam\itfam\tenit} \textfont\itfam=\tenit
\def\sl{\fam\slfam\tensl} \textfont\slfam=\tensl
\def\bf{\fam\bffam\tenbf} \textfont\bffam=\tenbf
\scriptfont\bffam=\sevenbf \scriptscriptfont\bffam=\fivebf
\def\tt{\fam\ttfam\tentt} \textfont\ttfam=\tentt
\def\ssr{\fam\ssrfam\tenssr} \textfont\ssrfam=\tenssr
\scriptfont\ssrfam=\sevenssr \scriptscriptfont\ssrfam=\fivessr
\def\mbi{\fam\mbifam\tenmbi} \textfont\mbifam=\tenmbi
\scriptfont\mbifam=\sevenmbi \scriptscriptfont\mbifam=\fivembi
\tt \ttglue=.5em plus .25em minus .15em
\normalbaselineskip=12pt
\bigskipamount=12pt plus4pt minus4pt
\medskipamount=6pt plus2pt minus2pt
\abovedisplayskip=12pt plus 3pt minus 9pt
\belowdisplayskip=12pt plus 3pt minus 9pt
\abovedisplayshortskip=0pt plus 3pt
\belowdisplayshortskip=7pt plus 3pt minus 4pt
\parskip=0.0pt plus 1.0pt
\eightfoot
\setbox\strutbox=\hbox{\vrule height8.5pt depth3.5pt width0pt}
\let\sc=\eightrm
\let\big=\tenbig \normalbaselines\rm}
\def\tenbig#1{{\hbox{$\left#1\vbox to8.5pt{}\right.\n@space$}}
\def\square{\mathord{\dalemb{4.9}{5}\hbox{\hskip1pt}}}}

\def\eightpoint{\def\rm{\fam0\eightrm}
\textfont0=\eightrm \scriptfont0=\sixrm \scriptscriptfont0=\fiverm
\textfont1=\eighti \scriptfont1=\sixi \scriptscriptfont1=\fivei
\textfont2=\eightsy \scriptfont2=\sixsy \scriptscriptfont2=\fivesy
\textfont3=\eightex \scriptfont3=\eightex \scriptscriptfont3=\eightex
\def\it{\fam\itfam\eightit} \textfont\itfam=\eightit
\def\sl{\fam\slfam\eightsl} \textfont\slfam=\eightsl
\def\bf{\fam\bffam\eightbf} \textfont\bffam=\eightbf
\scriptfont\bffam=\sixbf \scriptscriptfont\bffam=\fivebf
\def\tt{\fam\ttfam\eighttt} \textfont\ttfam=\eighttt
\def\ssr{\fam\ssrfam\eightssr} \textfont\ssrfam=\eightssr
\scriptfont\ssrfam=\sixssr \scriptscriptfont\ssrfam=\fivessr
\def\mbi{\fam\mbifam\eightmbi} \textfont\mbifam=\eightmbi
\scriptfont\mbifam=\sixmbi \scriptscriptfont\mbifam=\fivembi
\tt \ttglue=.5em plus .25em minus .15em
\normalbaselineskip=9pt
\bigskipamount=9pt plus3pt minus3pt
\medskipamount=5pt plus2pt minus2pt
\abovedisplayskip=9pt plus 3pt minus 9pt
\belowdisplayskip=9pt plus 3pt minus 9pt
\abovedisplayshortskip=0pt plus 3pt
\belowdisplayshortskip=5pt plus 3pt minus 4pt
\parskip=0.0pt plus 1.0pt
\setbox\strutbox=\hbox{\vrule height8.5pt depth3.5pt width0pt}
\let\sc=\sixrm
\let\big=\eightbig \normalbaselines\rm}
\def\eightbig#1{{\hbox{$\left#1\vbox to6.5pt{}\right.\n@space$}}
\def\square{\mathord{\dalemb{3.9}{4}\hbox{\hskip1pt}}}}

\def\vfootnote#1{\insert\footins\bgroup\footsuite
    \interlinepenalty=\interfootnotelinepenalty
    \splittopskip=\ht\strutbox
    \splitmaxdepth=\dp\strutbox \floatingpenalty=20000
    \leftskip=0pt \rightskip=0pt \spaceskip=0pt \xspaceskip=0pt
    \textindent{#1}\footstrut\futurelet\next\fo@t}
\def\hangfootnote#1{\edef\@sf{\spacefactor\the\spacefactor}#1\@sf
    \insert\footins\bgroup\footsuite
    \let\par=\endgraf
    \interlinepenalty=\interfootnotelinepenalty
    \splittopskip=\ht\strutbox
    \splitmaxdepth=\dp\strutbox \floatingpenalty=20000
    \leftskip=0pt \rightskip=0pt \spaceskip=0pt \xspaceskip=0pt
    \smallskip\item{#1}\bgroup\strut\aftergroup\@foot\let\next}
\def\footsuite{}

\def\tenfoot{\def\footsuite{\tenpoint}}
\def\eightfoot{\def\footsuite{\eightpoint}}
\catcode`@=12
\twelvepoint
\openup 1\jot
\oneandahalfspace
\def\crampest{\medmuskip = 1mu plus 1mu minus 1mu}
\def\uncramp{\medmuskip = 4mu plus 2mu minus 4mu}
\def\underbuildrel#1\under#2{\mathrel{\mathop{\kern0pt #2}\limits_{#1}}}
\def\subfor{\lower3pt\hbox{$|$}}
\def\gtlt{\mathrel{\raise4.5pt\hbox{\oalign{$\scriptstyle>$\crcr$\scriptstyle<$}}}}
\def\im{{\rm i}}
\def\R{\rlap I\mkern3mu{\rm R}}
\def\C{\mkern1mu\raise2.2pt\hbox{$\scriptscriptstyle|$}\mkern-7mu{\rm C}}
\def\arcsinh{\mathop{\rm arcsinh}\nolimits}
\nofirstpagenotwelve
\rightline{hep-th/9412168}
\rightline{Imperial/TP/94-95/10}
\vskip 1cm
\centerline{\bf Non-critical $d=2$ Gravities and Integrable
Models\hangfootnote{$^\ast$}{Contribution to the Proceedings
of the $6^{\rm th}$ Regional Conference on Mathematical Physics, Islamabad,
Pakistan, February 1994.}}
\vskip 1cm
\centerline{K.S.\ Stelle}
\bigskip
\centerline{\it The Blackett Laboratory}
\centerline{\it Imperial College}
\centerline{\it Prince Consort Road}
\centerline{\it London SW7 2BZ}
\vskip .5cm
\AB\singlespace
We review the origin of anomaly-induced dynamics in theories of $d=2$ gravity
from a BRST viewpoint and show how quantum canonical transformations may be
used
to solve the resulting Liouville or Toda models for the anomalous modes.
\AE\oneandahalfspace

     Two-dimensional worldsheet gravity models coupled to non-critical matter
systems provide a very useful workshop for investigating the way in which
induced dynamics can arise from anomalies. In this article, we shall first
review the way in which such anomalous dynamics can arise within the context of
BRST quantization. Then we shall present a technique for solving the resulting
anomalous quantum system by canonical transformations, implemented by
intertwining operators. We shall use these techniques to find the wavefunctions
for the minisuperspace limits of Liouville and Toda $d=2$ gravities. The
integrable-model developments discussed in this article are adapted from Ref.\
[1].

     We start from the action for a set of $d=2$ scalar fields $x^a$,
$a=1,\ldots,D$, coupled to worldsheet gravity,
$$ I=-\ft12\int d^2\sigma\sqrt{-\gamma}\gamma^{ij}\partial_i x^a\partial_j x^b
\eta_{ab}.\eqno(1)
$$ We pick a worldsheet parametrization, using light-cone indices
$i,j=+,-$,
$$
\gamma_{ij}=e^\omega\pmatrix{\tilde h&1\cr 1&h\cr},\eqno(2)
$$ so that for the contravariant metric density one has
$$
\sqrt{-\gamma}\gamma^{ij}=(1-h\tilde h)^{-\ft12}\pmatrix{-h&1\cr 1&-\tilde h},
\eqno(3)
$$ and the Weyl invariance of classical two-dimensional gravity is expressed by
the fact that the conformal factor $e^\omega$ drops out in (3). This invariance
is of course subject to anomalies, which we next shall discuss using the BRST
formalism.
\bigskip
\leftline{\it Anomalous Dynamics}

     In discussing the anomalies, we shall treat $\sigma^-=\sigma-\tau$ as the
``evolution'' coordinate on the worldsheet. Admittedly, this overlooks the fact
that the surfaces $\sigma^-={\rm const.}$ are not actually proper Cauchy
surfaces, because there exist some physical trajectories that do not cross
them,
but we shall not be concerned with this subtlety here. We shall, on the other
hand, be more careful with the gauge fixing. Treating
$\sigma^-$ as the evolution coordinate and $\sigma^+$ as a ``spatial''
coordinate means that the gauge symmetry for $h$, $\delta h=\partial_-k
+\ldots$, is similar to the transformation of the time component $A_0$ of the
Maxwell gauge field, and requires a derivative gauge-fixing term [2], imposed
by
a Lagrange multiplier, $\int \pi\partial_-h$. On the other hand, we shall treat
the left-moving sector gauge field $\tilde h$ as analogous to $A_3$ in Maxwell
theory, so we shall consider the gauge-fixing term
$\tilde\pi\tilde h$ to be acceptable. Similarly, we shall need to impose
gauge-fixing for the Weyl symmetry, using the gauge-fixing term
$\pi_\omega\omega$. When these gauge conditions are all satisfied, the
world-sheet metric takes the ``chiral light-cone gauge'' form
$$
\gamma_{ij}=\pmatrix{0&1\cr1&h\cr},\eqno(4)
$$ giving the simple form $-\int d^2\sigma[(\partial_+x^a\partial_-x^b-
h\partial_+x^a\partial_+x^b)\eta_{ab}$ for the scalar-field action.
Corresponding
to these gauge-fixing conditions, we shall need to introduce (antighost,ghost)
pairs $(b,c)$, $(\tilde b,\tilde c)$, $(b_\omega,c_\omega)$. The corresponding
gauge-fixed action taken together with the ghost action, which we shall denote
by
$S$, then has a tree-level BRST residuum of the original gauge symmetry,
$$\eqalignno{
\delta x^a&=c\partial_+x^a+\tilde c\partial_-x^a,\cr
\delta c=c\partial_+c+&\tilde c\partial_-c,\qquad
\delta\tilde c=\tilde c\partial_-\tilde c+c\partial_+\tilde c,\cr
\delta c_\omega&=c\partial_+c_\omega+\tilde c\partial_-c_\omega,&(5)\cr}
$$ etc.  This tree-level BRST symmetry can be encoded in the standard way [3]
by
including sources for the BRST variations of all fields,
$$\eqalignno{
\Sigma&=S+\int K_{q^A}\delta q^A\cr &=S+\int K_{x^a}(c\partial_+x^a+\tilde
c\partial_-x^a)+ K_{C}(c\partial_+c+\tilde c\partial_-c)\ +\ldots,&(6)\cr}
$$ where the generalized index $A$ runs over all of the fields of the theory,
including the ghosts and antighosts. Including sources for the variations in
this way allows us to write the tree-level BRST invariance simply as
$$ (\Sigma,\Sigma)=0,\eqno(7)
$$ where the Batalin-Vilkovisky antibracket [4] is defined by
$$ (A,B)=\int{\delta A\over\delta q^A}{\delta B\over\delta K_{q^A}}+ {\delta
A\over\delta K_{q^A}}{\delta B\over\delta q^A}.\eqno(8)
$$ Note that the ghost number of the $(A,B)$ antibracket is one more than the
sum
of the ghost numbers of $A$ and $B$, where ghost number is defined by
$G(c,\tilde c,c_\omega)=1$, $G(b,\tilde b,b_\omega)=-1$.
     Upon quantization, the extended classical action $\Sigma$ becomes the
tree-level limit of the quantum effective action
$\Gamma=\hbar^0\Sigma+\hbar^1\Gamma^{(1)}+\hbar^2\Gamma^{(2)}+\ldots$. If the
BRST symmetry (5) were unbroken at the full quantum level, one would expect to
have the quantum Ward identity $(\Gamma, \Gamma)=0.$ However, this identity is
disturbed by the presence of anomalies, giving instead the anomalous Ward
identity
$$ (\Gamma,\Gamma)=\Delta.\eqno(9)
$$ The anomaly $\Delta$ on the right-hand side of Eq.\ (9) is a local
expression
at lowest order, and at higher orders one encounters ``dressings'' of the
anomalies that have already occurred at lower orders, owing to their presence
in
subdiagrams. These dressings constitute the expected quantum corrections to
Green functions with an operator insertion. To express this more precisely, one
should really include a source $K_\Delta$ for every local expression $\Delta$
occurring in (9) and write the right-hand side of (9) as $\delta\Gamma/\delta
K_\Delta\equiv
\Delta\cdot\Gamma$. At order $\hbar^n$, if one subtracts out the anticipated
nonlocal dressings of lower-order anomalies, the residual anomalous terms of
order
$\hbar^n$ will be local. For our present purposes, it is sufficient to note
that
the structure of Eq.\ (9), together with the locality condition, yields the
Wess-Zumino consistency condition on the anomalies. At the one-loop order, use
of
Eqs (7) and (8) together with the Jacobi identity
$(A,(B,C))+(B,(C,A))+(C,(A,B))\equiv 0$ yields the consistency condition
$$ (\Sigma,\Delta)=0. \eqno(10)
$$

     Note that this condition is the same as that which governs the structure
of
renormalization counterterms, but here it is considered at ghost number one
instead of zero. In fact, for the anomalies we are strictly interested in those
solutions of (10) that {\it cannot} be removed by renormalization; the true
anomalies are solutions of (10) such that
$$
\Delta\ne(\Sigma,Y)\eqno(11)
$$ for any local functional $Y$ of ghost number zero. Eqs (10) and (11)
constitute the familiar cohomology problem of identifying the potential
anomalies for a theory. Anomaly functionals are representatives of cohomology
classes and may be changed in form by the addition of counterterms to the
action, but given (11) they cannot be completely removed. Once an anomaly has
occurred in a theory, however, counterterms noninvariant under its
corresponding
symmetry may then occur. Thus, anomalies and their associated induced
renormalizations need to be considered together.

     In the present case, the solutions to the consistency condition may be
written, after setting the variation sources $K_{q^A}$ to zero,
$$
\Delta=\alpha\int d^2\sigma \partial_+^2 hc_\omega + \beta\int d^2\sigma
c_\omega.\eqno(12)
$$ Only the $\alpha$ coefficient is strictly an anomaly in the sense of Eq.\
(11), {\it i.e.,} of not being removable by renormalization. However, once the
$\alpha$ coefficient is nonzero, the Weyl symmetry preventing the occurrence of
the $\beta$ coefficient as a renormalization counterterm is absent. The $\beta$
coefficient then arises as a finite residuum after renormalization. The
coefficients $\alpha$ and $\beta$ of the potential anomalies in (12) depend on
the central charge of the scalar fields $x^a$, and take the values
$\alpha=(c-26)/24$, $\beta=\mu_0^2(c-2)/24$, where $\mu_0$ is an infrared
regularization mass [5].  Further finite counterterms could be added to shift
the
anomaly away from the Weyl symmetry (with ghost $c_w$) into the general
coordinate symmetries ($c$, $\tilde c$), but we shall find it more appropriate
to leave them in the form (12).

     Given the presence of BRST anomalies in a theory, one may adopt one of two
approaches to studying the resulting dynamics. One may proceed by a direct
consideration of the correlation functions implied by the anomalous Ward
identity (9). Within the context of BRST quantization, such a direct approach
to
the anomalies has not been widely adopted. However, a related procedure of
holding off from integrating over the gauge fields in the generating-functional
path integral until after the anomalies have appeared from the integrals over
scalar ``matter'' fields has been used to show the existence of a hidden
$SL(2,\R)$ symmetry in the case of anomalous Liouville gravity [6]. A more
frequently-adopted way to extract the anomalous dynamics is to eliminate the
anomalies by compensation. In this procedure, an extra field is introduced into
the theory, frequently in the context of a symmetry-preserving regularization.
Classically, this extra field drops out of the theory by virtue of the as yet
unbroken gauge symmetries, but in the regularized theory this field has
non-trivial couplings and this fact allows for the possibility of a residual
non-trivial coupling for it in the renormalized quantum theory. It should be
emphasized that in such a compensation procedure, no anomalies actually occur
in
the BRST symmetry because the extra field allows for a removal of the potential
anomalies by finite renormalizations. But the finite renormalizations have the
effect of stopping the compensator from decoupling as it did at the classical
level. The anomalous dynamics is then the dynamics of this extra field.

     In the case of $d=2$ gravity theories with the conformal anomalies (12),
for
a compensation procedure one introduces a classically-decoupling scalar mode
$\phi(\sigma^+,\sigma^-)$ as an extra ``conformal'' mode for the
two-dimensional
metric, writing
$$
\hat\gamma_{ij}=e^{2\phi}\gamma_{ij},\eqno(13)
$$ and then rewriting the action with the replacement of $\gamma_{ij}$ by
$\hat\gamma_{ij}$. Of course, the $\phi$ field classically drops out owing to
the
original Weyl invariance of the action (1). Including $\phi$ into the formalism
produces the BRST transformations
$$\eqalignno{
\delta\phi=\delta_{\rm g.c.}+&\delta_\omega\phi;\cr
\delta_{\rm g.c.}\phi=c\partial_+\phi+\tilde c\partial_-\phi\quad &\quad
\delta_\omega\phi=-\ft12 c_\omega.&(14)\cr}
$$

     Writing the anomaly (12) in a manifestly generally-covariant fashion, one
has
$$
\Delta=\ft1{24}\int
d^2\sigma\sqrt{-\hat\gamma}[(c-26)R(\hat\gamma)+(c-2)\mu_0^2]c_\omega.\eqno(15)
$$ One may verify that this manifestly-covariant form reduces to the form (12)
upon use of the gauge-fixed form of the metric (4). The presence of the extra
field
$\phi$ changes our earlier discussion, however, in that (15) is no longer
cohomologically non-trivial and can be eliminated by a finite counterterm in
the
action. To see this, we note a lemma that holds in $d=2$,
$$
\sqrt{-\hat\gamma}R(\hat\gamma)=\sqrt{-\gamma}R(\gamma)-2\sqrt{-\gamma}
\nabla^2\phi,\eqno(16)
$$ so that
$$
\Delta=-\ft1{12}\int d^2\sigma[(c-26)(R(\gamma)-2\nabla^2\phi)
+(c-2)\mu_0^2e^{2\phi}]\delta_\omega\phi.\eqno(17)
$$ Thus, with $\phi$ included, $\Delta$ is now cohomologically trivial because
it may be eliminated by the finite local counterterm
$$
\delta\Sigma=\ft1{12}\int
d^2\sigma\sqrt{-\gamma}[(c-26)(\gamma^{ij}\partial_i\phi\partial_j\phi+
R(\gamma)\phi)+\ft12(c-2)\mu_0^2e^{2\phi}].\eqno(18)
$$ Note that by a redefinition of the $\phi$ field, $\phi\rightarrow\phi+{\rm
constant}$, one may alter the scale of the coefficient $\mu_0$. As a result,
the
specific value of this coefficient is not of physical importance; only the fact
that its value is nonvanishing is important. Although the counterterm (18)
removes the anomaly from the BRST symmetry, it violates the classical
decoupling
of the
$\phi$ field, which is now the expression of the anomaly in the compensated
formalism, and this gives rise to new dynamics not apparent in the original
classical theory. Note that for a subcritical matter central charge,
$c<26$, the action (18) for the scalar field $\phi$ is that for a positive-norm
mode.

     To continue further using the compensated-anomaly approach, one would like
to treat the anomaly-induced $\phi$ field from the start as an extra regular
scalar field in the action. This necessitates some changes to the above
picture.
For one thing, the kinetic term for $\phi$ in (18) does not have a standard
normalization. More serious is the fact that the functional measure for the
$\phi$ mode ({\it i.e.}\ the measure for the quantum-mechanical inner product)
is originally that for the conformal part of a worldsheet metric (13), and is
not the standard translation-invariant measure for an ordinary scalar field. It
is much more convenient for calculation to treat this field as an ordinary
scalar. Doing so causes it in turn to make a further contribution of $1$ to the
total central charge, arising from anomalous diagrams in which $\phi$ occurs in
loops. For our present purposes, we shall simply follow the argument of Ref.\
[7] and shall posit that the action for the $\phi$ mode is of the form (18),
but
with renormalized coefficients. Going over to a Euclidean signature for the
worldsheet now, we thus shall work with
$$ I_{\phi}=\int d^2z\sqrt\gamma(\ft12\gamma^{ij}\partial_i\phi\partial_j\phi +
QR(\gamma)\phi+\ft12\mu^2e^{\lambda\phi}),\eqno(19)
$$ where $z=e^{\tau+\im\sigma}$.

     The coefficients $Q$ and $\lambda$ need to be fixed by the requirement of
anomaly cancellation, for we shall still require the $\phi$ mode to eliminate
the anomalies by compensation. When it is treated as an ordinary field with a
translation-invariant functional metric, the $\phi$ field contributes an amount
$c_\phi=1+12Q^2$ to the central charge. Thus, the central charge condition for
anomaly cancellation after the change in the functional metric is [8]
$$ c_x+c_\phi=c_x+1+12Q^2=26,\eqno(20)
$$ fixing the value of the ``background charge'' coefficient $Q$ for a given
set
of ``matter'' scalar fields $x^a$. Similarly, requiring the cancellation of
anomalies arising from the presence of the potential term $e^{\lambda\phi}$,
which threatens to produce additional $\phi$-dependent anomalies beyond those
controlled by the central charge, one derives [9] that $\lambda$ take the value
$$
\lambda=Q-\sqrt{Q^2-2}=\ft1{12}(\sqrt{25-c}-\sqrt{1-c_x}).\eqno(21)
$$ This expression shows that imaginary values for the Liouville potential
coefficient occur for $c_x>1$, revealing one well-known aspect of the ``$c=1$
barrier,'' that is also clearly seen in the matrix model-treatment of
non-critical string theories. The proper handling of cases for $c_x>1$ remains
an important open problem.

     From the standpoint of conformal field theory, the anomaly cancellation
condition (21) may be understood as the requirement that the Liouville
potential
be an operator with (left,right) chiral weights equal to (1,1). The weight of
an
operator $e^{\alpha\phi}$ is obtained by taking an operator product of this
together with the holomorphic stress tensor
$$ T=T_{zz}=-\ft12(\partial\phi)^2-Q\partial^2\phi,\eqno(22)
$$ where the $Q$ term comes from the background charge term in the action (19).
The resulting operator-product expansion contains the singular terms
$$ T(z)e^{\alpha\phi(w,\bar w)}\sim {\Delta_\alpha e^{\alpha\phi(w,\bar
w)}\over
(z-w)^2} + {\partial(e^{\alpha\phi(w,\bar w)})\over (z-w)},\eqno(23)
$$ where $\Delta_\alpha=-\ft12 \alpha(\alpha+2Q)$, so that the value for
$\lambda$ selected in (21) gives $\Delta_\alpha=1$ as required.
\bigskip
\leftline{\it Higher-spin Worldsheet Symmetries}

     Two-dimensional theories admit a much greater variety of consistent gauge
symmetries than is possible in higher dimensions.  The gauge fields in
two-dimensional ``gravity'' theories all have enough local symmetries so that a
na\"\i ve count of their continuous degrees of freedom gives the result zero.
Correspondingly, there is no natural local candidate for a kinetic action for
the
spin-two gauge field, because the natural Einstein-Hilbert Lagrangian in $d=2$
becomes just the Euler-number density.  However, there is a natural
anomaly-induced dynamics for the $d=2$ metric, as we have seen. There is
nothing
in the derivation of this anomalous dynamics that restricts one to
consideration
of the spin-two metric, however. Indeed, it is now well-known that there exist
consistent closed quantum algebras with infinitely many different combinations
of
spin-two and higher-spin generators. When one works in the chiral light-cone
gauge (4) for the worldsheet metric, the residual chiral symmetry for the
spin-two gauge field $h$ is the Virasoro algebra. Extensions of the Virasoro
algebra that include higher-spin generators are known generically as $W$
algebras. Here, we shall be concerned principally with the most basic such
extension, the $W_3$ algebra [10],
$$
 \eqalignno{ T(z)T(w) &\sim {\partial T\over(z-w)} +
  {{2T}\over {(z-w)^2}} +{{ 1\over 2 } c\over(z-w)^4} &(24a) \cr T(z)W(w) &\sim
{{\partial W}\over {z-w}} + {3W\over(z-w)^2} &(24b) \cr W(z)W(w) &\sim
{1\over(z-w)} \Big( {1\over 15}\partial^3 T + {16\over 22+5c} {\partial
\Lambda}
\Big )
\cr &+ {1\over(z-w)^2} \Big({3\over 10}{\partial }^2 T + 2{16\over
22+5c}\Lambda
\Big ) \cr &+ {\partial T\over(z-w)^3} + {2T\over(z-w)^4} + {{1\over 3}
c\over(z-w)^6}, &(24c)
\cr }
$$ where $\Lambda$ is a composite operator,
$$
\Lambda(z) = :TT:(z)-{3\over 10} {\partial }^2 T(z),
\eqno(25)
$$ in which the colons denote normal ordering.

     The $W_3$ algebra has a central-charge structure that is determined by the
central charge in its Virasoro subalgebra. In order to realize local $W_3$
symmetry, gauge fields of spin two (the $h$ component of the usual metric in
the
gauge (4)) and spin three (generally denoted $B$) are needed. Upon gauge
fixing,
both of these symmetries require ghosts and both the spin-two and spin-three
ghosts contribute to the central charge that must be canceled by the
compensating
scalars $\phi_{1,2}$ and the ``matter'' scalar fields
$x^a$. In the $W_3$ case, the central charge that needs to be canceled is [11]
$c_{\rm gh}=-26-74=-100$, where the $-74$ contribution comes from the
spin-three
ghosts. It turns out that, given the requirement to cancel independent
anomalies
in both the spin-two and spin-three currents, there is no ``critical'' set of
free fields
$x^a$ for which all the anomalies cancel, not even for $100$ scalars. Thus, a
compensating-field mechanism similar to that discussed above in the Liouville
gravity case is essential. Realizations with arbitrary numbers of $x^a$ fields
exist [12]; the simplest realization is the original one [13], with no $x^a$
fields, but with two compensating fields $\phi_1$ and $\phi_2$. An anomaly-free
realization requires both of these fields to have background charges; their
chiral stress-tensor is just the sum
$$ T=T_1+T_2=[-\ft12(\partial\phi_1)^2-Q_1\partial^2\phi_1] +
[-\ft12(\partial\phi_2)^2-Q_2\partial^2\phi_2].\eqno(26)
$$ For this stress tensor, one has a central-charge contribution
$c=c_1+c_2=100$
for the values
$$ Q_1=\sqrt{{49\over 8}}\qquad\qquad Q_2=\sqrt{{49\over 24}};\eqno(27)
$$ this set leads to a cancellation of anomalies both in the spin-two
stress-tensor algebra and also in the algebra of the spin-three generator,
whose
tree-level limit is given by
$$ W=\ft13(\partial\phi_1)^3+Q_1\partial\phi_1\partial^2\phi_1 +
\ft13\partial^3\phi_1 + 2\partial\phi_1T_2 + Q_1\partial T_2.\eqno(28)
$$ Note that the second compensating field $\phi_2$ occurs in the spin-three
current only through its stress-tensor $T_2$. This feature persists when
quantum
corrections to the realization (26) are taken into account, so that one may
identify $\phi_2$ as the compensating field for the Virasoro subalgebra, while
$\phi_1$ is the compensator for the spin-three sector.

     As in the Liouville case, once the local $W_3$ symmetry has been broken by
anomalies, new counterterms become possible in the theory and after the
corresponding renormalizations, one should have finite residual interaction
terms generalizing the Liouville potential $e^{\lambda\phi}$. Following the
same
logic of demanding the cancellation of potential compensating-field-dependent
anomalies [9], or by demanding that the the corresponding operators be of
weight
(1,1) with respect to the full $W_3$ algebra [14], one obtains the allowed
generalizations of the Liouville potential,
$$\eqalignno{ V_1&=e^{-{3\over7}Q_1\phi_1+{3\over7}Q_2\phi_2}&(29a)\cr
V_2&=e^{-{6\over7}Q_2\phi_2}.&(29b)\cr}
$$ As in the Liouville case, the magnitudes of the coefficients of these
potentials may be altered by constant shifts of $\phi_{1,2}$, so the only
physically-meaningful aspect of the coefficients of these potentials is their
non-vanishing. The potentials (29) taken together with the kinetic terms for
$\phi_{1,2}$ describe an
$A_2$ Toda field theory.
\np
\leftline{\it Minisuperspace Approximation}

     In the presence of interaction potentials such as the Liouville potential
$e^{\lambda\phi}$ or the Toda potentials (29), the dynamics of the compensating
modes is discretely changed with respect to the dynamics of free fields, even
thought the Liouville and Toda theories are integrable field theories. We shall
review how some of these differences come about, concentrating on the
center-of-mass modes of the compensating fields, which are the ones most
affected
by the potentials. Inclusion of the oscillating-string modes may then be
carried
out consistently within the context of perturbation theory, after the
non-perturbative dynamics of the center-of-mass modes has been understood. The
separation of the center-of-mass modes is known as the ``minisuperspace
approximation,'' where reference is made to a subspace of the ``superspace''
configuration space of metric states, and not to supersymmetry. In the
Liouville
case, one splits up the field $\phi$ as follows:
$$
\phi(\tau,\sigma)={2\over\lambda}q(\tau)+\phi^{\rm osc}(\tau,\sigma),\eqno(30)
$$ where $\phi^{\rm osc}(\tau,\sigma)$ is defined to satisfy $\oint d\sigma
\phi^{\rm osc}(\tau,\sigma)=0$, and the coefficient in front of $q(\tau)$ is
for
convenience of normalization in the minisuperspace action. Recall that
$\lambda$
is given in terms of $Q$ by (21). (In most of the following, we shall
concentrate on the case of ``pure'' Liouville gravity, for which
${2\over\lambda}=-{6Q\over5}$). In the Toda case, the splitup is (specializing
to
the two-field pure Toda gravity case, for which the background charges are as
given in (27))
$$\eqalignno{
\phi_1(\tau,\sigma)&=-{4Q_1\over7}q_1(\tau)+\phi_1^{\rm
osc}(\tau,\sigma)&(31a)\cr
\phi_2(\tau,\sigma)&=-{4Q_2\over7}(2q_2(\tau)-q_1(\tau))+\phi_2^{\rm
osc}(\tau,\sigma),&(31b)\cr}
$$ where again $\oint d\sigma\phi_{1,2}^{\rm osc}(\tau,\sigma)=0$. In
extracting
the
$\sigma$-independent modes in (30,31), we are using the original
$(\tau,\sigma)$
coordinates of a cylindrical string worldsheet instead of the complex
$z=e^{\tau+\im\sigma}$ coordinates generally used in conformal field theory.
The
change from the coordinate $z$ to the coordinate $w=\ln z$ is effected by a
conformal transformation on the worldsheet, generated by the stress tensor
$T(z)$. Owing to the background-charge terms $Q$ in (22,24), the
transformations
of the compensating fields are not quite those of ordinary scalars, but instead
give, in the Liouville case,
$$
\partial_z\phi\rightarrow(\partial_w\phi-Q)z^{-1},\eqno(32)
$$ so that in the transition from $z$ to $w$ the momentum carried by a $\phi$
state is modified according to $p_\phi\rightarrow p_\phi(w)=p_\phi(z)-\im Q$.
This shift must be taken into account in comparing free-field states
constructed
using conformal-field theory with the minisuperspace wavefunctions that we
shall
discuss. Of course, the overall wavefunction in a string theory including the
compensating modes $\phi$ or $\phi_{1,2}$ will be subject to the constraints
following from varying $h$ and $B$ in the action. These constraints impose
``mass-shell'' conditions that include contributions from the background
charges. For our present purposes, however, we shall be content to treat such
questions at the string-theory level in a perturbative fashion once we have
understood the dynamics of the interacting compensating-field sectors in
isolation. In particular, we shall be interested in finding the wavefunctions
for
the minisuperspace modes $q$ and $q_{1,2}$.
\bigskip
\leftline{\it Canonical Transformations for Integrable Models}

      The Liouville and Toda systems that emerge as the Lagrangians of the
anomalous modes in ordinary and $W$-string theories are famous examples of
integrable systems. They are integrable at the classical level because they
possess sufficiently large symmetry algebras to give conserved quantities
corresponding to all the degrees of freedom. This does not guarantee that these
systems remain integrable at the quantum level, although this does in fact
prove
to be the case. Many different approaches have been followed in studying these
problems. Owing to the importance of vertex operators in string theory, much
effort has been expended on the promotion of exponentials of field operators to
their analogues at the quantum level, taking into account the requirements of
locality and covariance. One should mention along these lines especially the
work
of Gervais, Neveu and collaborators [15], together with the recent work of
[16].
Here, we shall follow a different line of attack in concentrating on the actual
wavefunctions of the Liouville and Toda theories. At the present stage of
development of this approach, we shall confine our attention to the
minisuperspace level discussed above. We shall aim to derive the wavefunctions
of
these theories by applying canonical transformations to map these interacting
models onto corresponding free-field theories. In the process, a characteristic
feature shall emerge: these canonical transformations are multi-valued (a
feature
also important in the approaches of [15,16]), so the relation to free-field
theories is modified by the need to take a quotient of these free theories by
the
Weyl groups of the corresponding interacting Toda systems. The Weyl group
symmetry plays a crucial r\^ole in the structure of the resulting integral
representations for the wavefunctions.

     We begin with the Liouville case. The classical Liouville Hamiltonian is
$$ H_{\rm L}=\ft12(p^2 + e^{2q}),\eqno(33)
$$ and in the following we shall let the evolution parameter be denoted by by
$\tau=t$. The equations of motion following from (33) are integrable, since for
the one variable $q$, we have a corresponding conserved quantity, namely
$H_{\rm L}$ itself. This equality of the numbers of conserved quantities and
independent variables persists also at the full field-theory level, owing to
the
infinite-dimensional Virasoro symmetry of the model. The general solution to
the
classical equations of motion following from (33) may be written
$$ q=-\ln\left({1\over\tilde p}\cosh(\tilde q(t)\right),\eqno(34)
$$ where $\tilde q=\tilde p(t-t_0)$, and $\tilde p$ and $t_0$ are the two
expected integration constants of the motion. Writing the general classical
solution in this way suggests the following canonical transformation, in which
$(\tilde q,\tilde p)$ are now interpreted as a new pair of phase-space
variables:
$$\eqalignno{ e^{-q}&={1\over\tilde p}\cosh\tilde q&(35a)\cr p&=-\tilde
p\tanh\tilde q.&(35b)\cr}
$$ The usefulness of this canonical transformation is that in the new $(\tilde
q,\tilde p)$ variables, the Hamiltonian becomes
$$
\tilde H_{\rm L}=\ft12 \tilde p^2,\eqno(36)
$$ which makes it plain that in the $(\tilde q,\tilde p)$ variables we have a
free system.

     An important feature of the transformation from (33) to (36) is that the
canonical transformation between them has a branch structure. The free
Hamiltonian (36) has reflection symmetry in momentum space, $\tilde
p\rightarrow-\tilde p$; this has the consequence that the inverse map from
$(\tilde q,\tilde p)$ to $(q,p)$ is {\it two-to-one}. Both free-variable
motions
$(\tilde q,\tilde p)$ and $(-\tilde q,-\tilde p)$ correspond to the same
solution
of the interacting system $(q,p)$. Clearly, for real $p$, Eq.\ (35$a$) cannot
be
solved for real $\tilde p<0$, but for that case there is another canonical
transformation that maps to a free system, obtained by flipping the signs of
$\tilde q$ and $\tilde p$ in (35). Consequently, the general transformation to
the free system can be written $e^{-q}={1\over|\tilde p|}\cosh\tilde q;\
p=-\tilde p\tanh\tilde q$, which makes the branch structure transparent. The
$Z_2$ transformation on the free variables can be identified with the Weyl
group
of the underlying $A_1=SL(2,\R)$ group of the Liouville theory.

     At the quantum level, one has to contend with the non-commuting nature of
field operators. Nonetheless, one can still find a canonical transformation at
the quantum level for the Liouville case if one first takes care to split up
the
overall transformation between the interacting and the free theories into small
substeps, each of which remains canonical even when taking account of operator
ordering and also which has a clear effect on quantum wavefunctions. Letting
the
overall generator of the transformation be denoted $C$, the canonical
transformation may be written
$$ CH_{\rm L}C^{-1}=\tilde H_{\rm L},\eqno(37)
$$
 from which it is clear that what we are looking for is an operator that {\it
intertwines} between the free and interacting Hamiltonians. The technique of
constructing canonical transformations as intertwining operators has been
developed by Anderson in Refs [17].

     In the Liouville case, one decomposes the transformation into the
following
sequence of subtransformations [1]:
$$
\matrix{
\rlap{[L1]}\hskip2cm&{\cal P}_{\ln q}: &\hskip1cm\rlap{$q\mapsto\ln q$}\hskip
5cm&\rlap{$p\mapsto qp$}\hskip2cm\cr
\rlap{[L2]}\hskip2cm&\cal I: &\hskip1cm\rlap{$q\mapsto p$}\hskip
5cm&\rlap{$p\mapsto -q$}\hskip2cm\cr
\rlap{[L3]}\hskip2cm&p^{-1}: &\hskip1cm\rlap{$q\mapsto p^{-1} q p=q+ \im
p^{-1}$}\hskip 5cm&\rlap{$p\mapsto p$}\hskip2cm\cr
\rlap{[L4]}\hskip2cm&{\cal P}_{\sinh q}: &\hskip1cm\rlap{$q\mapsto\sinh
q$}\hskip 5cm&\rlap{$p\mapsto{1\over\cosh q}p.$}\hskip2cm\cr}\eqno(38)
$$ It may be verified that each of the subtransformations in (38) is canonical
in the quantum sense of preserving the canonical commutation relation
$[p,q]=-\im$. Transformations [L1] and [L4] are point transformations, and have
a straightforward action on Schr\"odinger representation wavefunctions, for
$q\mapsto q'$, $\psi(q)\mapsto\psi(q')$. Transformation [L2] is implemented on
wavefunctions by a Fourier transformation. Transformation [L3] is implemented
on
Schr\"odinger representation wavefunctions by indefinite integration in the
argument $q$ and multiplication by $-\im$. The overall transformation after
combining [L1--L4] may be written
$$\eqalignno{ e^{-q}&={1\over\tilde p}\cosh\tilde q&(39a)\cr p&=-\tanh(\tilde
q)\tilde p,&(39b)\cr}
$$ showing that, remarkably, the quantum canonical transformation is actually
one of the simple ordering choices for the operators in (35). Corresponding to
[L1--L4], we have the sequence of transformed Hamiltonians:
$$
\eqalign{ 2H_{\rm L}&=p^2 + e^{2q}\cr [{\rm L1}]\qquad\qquad\qquad&\mapsto (q
p)^2 +q^2=q^2p^2-\im q p +q^2\cr [{\rm L2}]\qquad\qquad\qquad&\mapsto p^2 q^2
+\im p q +p^2\cr [{\rm L3}]\qquad\qquad\qquad&\mapsto p q^2 p +\im q p  +p^2
=(1+q^2) p^2 -\im q p=\Big[(1+q^2)^{\ft12} p\Big]^2\cr [{\rm
L4}]\qquad\qquad\qquad&= \tilde p^2 =2 \tilde H_{\rm L}.\cr}
\eqno(40)
$$

     The generator $C$ of the overall transformation (39) intertwines between
$H_{\rm L}$ and $\tilde H_{\rm L}$, as we have seen. Using $C^{-1}$, we may
obtain an eigenfunction of the interacting Hamiltonian by operating on a
free-Hamiltonian eigenfunction $\tilde\psi_k(\tilde q)=\exp(\im k\tilde q)$.
Since
$C$ intertwines between $H_{\rm L}$ and $\tilde H_{\rm L}$, the resulting
interacting-theory wavefunction must have the same eigenvalue, $\ft12 k^2$, as
for the free wavefunction. The inverse intertwining operator is, from (38),
$$ C^{-1}={\cal P}_{e^q}\,{\cal I}^{-1}\, p\, {\cal P}_{\arcsinh q}.\eqno(41)
$$ In this way, one obtains
$$\eqalignno{
\psi_k(q)&\sim{k\over\sqrt{2\pi}}\int_0^\infty dy\,e^{-\ft\im 2 e^q(y-y^{-1})}
y^{\im k-1}\cr &={2k\over\sqrt{2\pi}}e^{\pi k\over 2}K_{\im k}(e^q),&(42)\cr}
$$ where $K_{\im k}$ is a modified Bessel function. Now we have to face the
issue of normalization. The transformation (37) is canonical but is not
unitary.
As a consequence, normalization is not preserved; another way of expressing
this
is that the transformation has a non-trivial action also on the
quantum-mechanical inner product. In order to have a properly-normalized
Liouville wavefunction with respect to the standard quantum-mechanical inner
product, a normalization factor must be supplied. The final result, normalized
to a delta function
$\delta(k-k')$, is
$$
\psi_k(q)={1\over\pi}\sqrt{2k\sinh(\pi k)}\,K_{\im  k}(e^q).\eqno(43)
$$ In this result, this we note two related features. First, as a result of the
symmetry of the modified Bessel function in its $\im k$ index, the $Z_2$
Weyl-group symmetry is now manifest in the interacting Liouville wavefunction,
{\it i.e.}\ $\psi_k(q)=\psi_{-k}(q)$. Second, the zero-eigenvalue wavefunction
for
$k=0$, which was an acceptable delta-function normalizable wavefunction for the
free Hamiltonian $\tilde H_{\rm L}$, drops out of the normalizable spectrum for
the interacting Hamiltonian $H_{\rm L}$.

     Now consider the case of $W_3$ gravity. In the minisuperspace
approximation,
with the parametrization (31) for the center-of-mass modes, the Hamiltonian
becomes
$$  H_{\rm T}=\ft13(p_1^2+p_2^2+p_1p_2)+e^{2q_1-q_2}+e^{2q_2-q_1}.\eqno(44)
$$  In addition to the Hamiltonian, we have also the spin-three generator (28),
whose minisuperspace limit is
$$ W_{\rm T}=\ft1{18}(2p_1+p_2)(2p_2+p_1)(p_1-p_2) +\ft12 (2p_2+p_1)
e^{2q_1-q_2}-\ft12 (2p_1+p_2) e^{2q_2-q_1}.\eqno(45)
$$ The existence of these two first integrals and the consequent equality of
the
numbers of conservation laws and degrees of freedom makes Toda mechanics
classically integrable. Following the pattern of the Liouville discussion, the
classical solution leads to a canonical transformation over to free-field
phase-space variables
$(\tilde q_i,\tilde p_i)$, $i=1,2$:
$$
\eqalignno{ e^{-q_1}&={1\over \tilde p_1(\tilde p_1-\tilde p_2)}e^{\tilde q_1}
+
{1\over
\tilde p_2(\tilde p_1-\tilde p_2)}e^{\tilde q_2}+{1\over \tilde p_1\tilde
p_2}e^{-\tilde q_1-\tilde q_2}&(46a)\cr e^{-q_2}&={1\over \tilde p_1(\tilde
p_1-\tilde p_2)}e^{-\tilde q_1} + {1\over
\tilde p_2(\tilde p_1-\tilde p_2)}e^{-\tilde q_2}+{1\over \tilde p_1\tilde
p_2}e^{\tilde q_1+\tilde q_2}&(46b)\cr (2p_1+p_2)e^{-q_1}&=-{(2\tilde
p_1-\tilde
p_2)\over \tilde p_1(\tilde p_1-\tilde p_2)}e^{\tilde q_1} - {(2\tilde
p_2-\tilde p_1)\over \tilde p_2(\tilde p_1-\tilde p_2)}e^{\tilde q_2}+{(\tilde
p_1+\tilde p_2)\over \tilde p_1\tilde p_2}e^{-\tilde q_1-\tilde q_2}&(46c)\cr
(2p_2+p_1)e^{-q_2}&={(2\tilde p_1-\tilde p_2)\over \tilde p_1(\tilde p_1-\tilde
p_2)}e^{-\tilde q_1} + {(2\tilde p_2-\tilde p_1)\over \tilde p_2(\tilde
p_1-\tilde p_2)}e^{-\tilde q_2}-{(\tilde p_1+\tilde p_2)\over \tilde p_1\tilde
p_2}e^{\tilde q_1+\tilde q_2}.&(46d)\cr}
$$

     The transformations (46) yield a free Hamiltonian and also a purely cubic
version of the spin-three conserved quantity (45):
$$\eqalignno{
\tilde H_{\rm T}&=\ft13(\tilde p_1^2+\tilde p_2^2-\tilde p_1\tilde p_2)&(47)\cr
\tilde W_{\rm T}&=\ft1{18}(2\tilde p_1-\tilde p_2)(2\tilde p_2-\tilde
p_1)(\tilde p_1+ \tilde p_2).&(48)\cr}
$$  As in the Liouville case, the map between the interacting and free theories
has a branch structure, now described by the Weyl group for the $A_2$ Toda
theory, which is a symmetry of the free-theory invariants (47,48). In this
case,
the Weyl group is the discrete group $S_3$, whose six elements are generated by
a
threefold rotation
$$ M:\qquad (\tilde q_1,\tilde q_2;\tilde p_1,\tilde p_2)\rightarrow (-\tilde
q_1-\tilde  q_2,\tilde q_1;-\tilde p_2,\tilde p_1-\tilde p_2),\eqno(49)
$$  and a twofold reflection
$$ R:\qquad (\tilde q_1,\tilde q_2;\tilde p_1,\tilde p_2)\rightarrow (\tilde
q_2,\tilde q_1;\tilde p_2,\tilde p_1).\eqno(50)
$$ As a result, the map from the free variables $(\tilde q_i,\tilde p_i)$ to
the
interacting variables $(q_i,p_i)$ is six-to-one. Just as in the Liouville case,
where all the {\it distinct} motions in the interacting theory are obtained
from
momenta $\tilde p>0$, so in the Toda case all the distinct motions of the
interacting theory are obtained by mapping from free-theory momenta that lie in
the {\it principle Weyl chamber}: $\tilde p_1>\tilde p_2>0$.

     Once again, it turns out to be possible to promote classical integrability
into quantum integrability by factorizing the canonical transformation (46)
into
a sequence of subtransformations, each of which has a clear effect on
wavefunctions [1]:
\crampest
$$
\matrix{\rlap{[T1]}\hskip.5cm&{e^{\ft\pi2(p_1+p_2)}\over\Gamma(1-\im(p_1+p_2))}:
&\rlap{$\left\{
\matrix{e^{q_1}\mapsto -e^{q_1}(p_1+p_2),\cr e^{q_2}\mapsto
-e^{q_2}(p_1+p_2),\cr}\right.$}\hskip 4.6cm&\rlap{$\matrix{p_1\mapsto p_1\cr
p_2\mapsto p_2\cr}$}\hskip 5.75cm\cr\strut\cr
\rlap{[T2]}\hskip.5cm&{\cal P}_{(\ln q_1,\ln q_2)}:&\rlap{$\left\{
\matrix{q_1\mapsto\ln q_1,\cr q_2\mapsto\ln q_2,\cr}\right.$}
\hskip 4.6cm&\rlap{$\matrix{ p_1\mapsto q_1p_1\cr p_2\mapsto q_2p_2\cr}$}
\hskip 5.75cm\cr\strut\cr
\rlap{[T3]}\hskip.5cm&q_1^{-1}q_2^{-2}:&\rlap{$\left\{
\matrix{q_1\mapsto q_1,\cr q_2\mapsto q_2,\cr}\right.$}\hskip 4.6cm&
\rlap{$\matrix{p_1\mapsto p_1-{\im\over q_1}\cr  p_2\mapsto p_2-{\im\over
q_2}\cr}$}\hskip 5.75cm\cr\strut\cr
\rlap{[T4]}\hskip.5cm &\exp\left(-\im({q_1^2\over q_2} + {q_2^2\over
q_1})\right):&\rlap{$\left\{
\matrix{q_1\mapsto q_1 ,\cr q_2\mapsto q_2,\cr}\right.$}
\hskip 4.6cm&\rlap{$\matrix{p_1\mapsto p_1-{q_2^2\over q_1^2}+{2q_1\over q_2}
\cr p_2\mapsto p_2-{q_1^2\over q_2^2}+{2q_2\over q_1}\cr}$}
\hskip 5.75cm\cr\strut\cr
\rlap{[T5]}\hskip.5cm&{\cal I}_1{\cal I}_2:&\rlap{$\left\{
\matrix{q_1\mapsto p_1,\cr q_2\mapsto p_2,\cr}\right.$}
\hskip 4.6cm&\rlap{$\matrix{p_1\mapsto -q_1\cr p_2\mapsto -q_2\cr}$}
\hskip 5.75cm\cr\strut\cr
\rlap{[T6]}\hskip.5cm&\rlap{${\cal P}_{(q_1-q_2+{1\over q_1q_2},{1\over
q_1}-{1\over q_2}+q_1q_2)}:$}\hskip4.25cm&\rlap{$\left\{
\matrix{q_1\mapsto q_1'=\phantom{\rlap{${1\over\det{\partial q_i'\over
\partial q_j}}$}}q_1-q_2+{1\over q_1q_2},\cr q_2\mapsto
q_2'=\phantom{\rlap{${1\over\det{\partial q_i'\over\partial q_j}}$}}{1\over
q_1}-{1\over q_2}+q_1q_2,\cr}\right.$}
\hskip 4.6cm&\rlap{$\matrix{p_1\mapsto p_1'={1\over\det{\partial
q_i'\over\partial q_j}}\left({\partial q_2'\over\partial q_2}p_1- {\partial
q_2'\over\partial q_1}p_2\right)\cr p_2\mapsto p_2'={1\over\det{\partial
q_i'\over\partial q_j}}\left({\partial q_1'\over\partial q_1}p_2-{\partial
q_1'\over\partial q_2}p_1\right)\cr}$}
\hskip 5.75cm\cr\strut\cr
\rlap{[T7]}\hskip.5cm&{\cal P}_{(e^{q_1},e^{q_2})}:&\rlap{$\left\{
\matrix{q_1\mapsto e^{q_1},\cr q_2\mapsto e^{q_2},\cr}\right.$}
\hskip 4.6cm&\rlap{$\matrix{p_1\mapsto e^{-q_1}p_1\ \cr p_2\mapsto
e^{-q_2}p_2.\cr}$}\hskip 5.75cm\cr}\eqno(51)
$$
\uncramp Among the transformations composing this free-field map, we have a
type
not yet encountered, the ``similarity'' transformations embodied in [T1,T3,T4]
(although, strictly speaking, the inverse-momentum transformations [L3] are
also
of this type). The coordinate similarity  transformations [T3,T4], of the form
($p_i\mapsto p_i-f,_i(q_j)$,
$q_i\mapsto q_i$) are generated by $e^{\im f(q_j)}$, transforming wavefunctions
$\Psi(q_j)$ into $e^{\im f(q_j)}\Psi(q_j)$ [17]. Momentum versions such as
[T1],
of the form  ($q_i\mapsto q_i+f,_i(p_j)$, $p_i\mapsto p_i$), are generated by
$e^{\im f(p_j)}={\cal I} e^{\im f(q_j)} {\cal I}^{-1}$. The sequence of steps
evolving the interacting into the free Hamiltonian is
$$
\eqalign{ 3 H_{\rm T}&=p_1^2+p_2^2+p_1p_2+3e^{2q_1-q_2}+3e^{2q_2-q_1}\cr [{\rm
T1}]\qquad\qquad&\mapsto p_1^2+p_2^2+p_1p_2-3(e^{2q_1-q_2}+
e^{2q_2-q_1})(p_1+p_2)\cr [{\rm T2}]\qquad\qquad&\mapsto
(q_1p_1)^2+(q_2p_2)^2+q_1p_1q_2p_2- 3\left({q_1^2\over q_2}+{q_2^2\over
q_1}\right)(q_1p_1+q_2p_2)\cr [{\rm T3}]\qquad\qquad&\mapsto
(p_1q_1)^2+(p_2q_2)^2+p_1q_1p_2q_2- 3\left({q_1^2\over q_2}+{q_2^2\over
q_1}\right)(p_1q_1+p_2q_2)\cr [{\rm T4}]\qquad\qquad&\mapsto
(p_1q_1)^2+(p_2q_2)^2+p_1q_1p_2q_2 -9q_1q_2-3p_2q_1^2-3p_1q_2^2\cr [{\rm
T5}]\qquad\qquad&\mapsto (q_1p_1)^2+(q_2p_2)^2+q_1p_1q_2p_2
-9p_1p_2+3q_2p_1^2+3q_1p_2^2\cr [{\rm T6}]\qquad\qquad&\mapsto
(q_1p_1)^2+(q_2p_2)^2-q_1p_1q_2p_2\cr [{\rm T7}]\qquad\qquad&\mapsto
p_1^2+p_2^2-p_1p_2=3\tilde H_{\rm T}.\cr}\eqno(52)
$$

     The generator $C$ of the transformation (51), which intertwines between
the
interacting and free theories to yield $C H_{\rm T} C^{-1}=\tilde H_{\rm T}$
and
$C W_{\rm T} C^{-1}=\tilde W_{\rm T}$, also gives the Toda wavefunction by
acting
on a free wavefunction, $\Psi_{k_1,k_2}(q_1,q_2)\sim
C^{-1}e^{\im(k_1q_1+k_2q_2)}$, with the result
$$
\eqalign{
\Psi_{k_1,k_2}(q_1,q_2)&={N_{k_1k_2}\over2\pi}e^{\pi k_1}\int_0^\infty du\,
e^{q_1+q_2}u^{-2}e^{-u-(e^{2q_1-q_2}+e^{2q_2-q_1})u^{-1}}\times\cr
&\qquad\int_0^\infty dy_1 \int_0^{\infty} dy_2
\,[{\rm jac}]\, y_1^{\im k_1} y_2^{\im k_2}  e^{ -e^{q_1}(y_1+y_2+{1\over
y_1y_2})u^{-1}} e^{ -e^{q_2}({1\over y_1}+{1\over y_2}+y_1y_2)u^{-1}},\cr
}\eqno(53)
$$ where
$$ [{\rm jac}]={1\over y_1 y_2}(y_1-y_2) (y_2-{1\over y_1^2}) (y_1-{1\over
y_2^2})\eqno(54)
$$ and $N_{k_1k_2}$ is a normalization factor. This result is manifestly
convergent and falls away quickly under the Toda potential, so that the
normalization factor
$N_{k_1k_2}$ is calculable as a convergent integral obtained using (53). As in
the Liouville case, the result after normalization should be fully Weyl-group
symmetric, but the zero momentum state $(0,0)$ is again not normalizable and so
drops out of the spectrum. Thus Toda theory is also a theory without a vacuum
state. The result (53) for the Toda wavefunction is of a different form from
previous results obtained principally by reduction of wavefunctions on group
manifolds [18], but these earlier forms may also be obtained by modifications
of
the intertwining-operator procedure [1].
\bigskip
\leftline{\it Vertex Operators versus States}

     Now let us return to one of the underlying issues of conformal field
theory
and of string theory, namely the relation between vertex operators and states,
using the insights gained from the above canonical transformations. This
relation is clear enough in the case of free-field theory, but it is worth
re-examining carefully in the more complicated cases with interacting Liouville
or Toda modes. The link between an operator $\cal O$ and its associated state
$\psi_{\cal O}\big(\phi(\sigma)\big)$ at some time $\tau$ is frequently written
in string theory as a path integral,
$$
\psi_{\cal O}\big(\phi(\sigma)\big)=\underbuildrel{{\cal
D};\,\xi\subfor_{\partial{\cal
D}}=\phi(\sigma)}\under{\int[d\xi(\sigma_i)]}e^{-\im I_{\rm L}}\,{\cal
O}(\xi),\eqno(55)
$$ where the point on the worldsheet at which $\cal O$ acts locally is taken to
correspond to negative temporal infinity $\tau\rightarrow -\infty$, and the
domain of integration $\cal D$ for the $[d\xi]$ integral is over all
worldsheets
bounded by an end loop $\partial{\cal D}$ corresponding to the evaluation time
$\tau$, on which Dirichlet boundary conditions
$\xi\subfor_{\partial{\cal D}}=\phi(\sigma)$ are imposed. In free-field theory,
which has a Fock-space interpretation and a normalizable vacuum state
$|0\rangle$, this reproduces the usual conformal-field-theory expression
$|{\cal
O}\rangle  =\lim_{z\rightarrow 0}{\cal O}(\phi(z))|0\rangle$ for the state
associated to $\cal O$.

     In our interacting theories, we may use our canonical transformations to
evaluate path integrals such as (55). We shall consider the Liouville state
associated to a vertex operator ${\cal O}=e^{\alpha\phi(z)}$, but shall
restrict
our discussion to the minisuperspace limit $\phi(z)\rightarrow q(t)$ and to the
tree level. The expression for $\psi_{\cal O}(q)$ becomes just the
path-integral
form of the evolution operator from $t_0$ to $t$ applied to an initial
wavefunction
${\cal O}(q(t_0))$ where $t_0\rightarrow -\infty$,
$$
\psi_{\cal O}\big(\phi(t)\big)=\lim_{t_0\rightarrow-\infty}e^{-\im H_{\rm
L}(t-t_0)}{\cal O}\big(\phi(t_0)\big).\eqno(56)
$$ Letting $\alpha=\im p$, so ${\cal O}_p=e^{\im pq}$, means starting off at
$t_0\rightarrow -\infty$ with a simple plane wave even though this is
definitely
not an eigenstate (43) of the Liouville theory. Most non-stationary state
wavefunctions dissipate in quantum mechanics, so it requires special
circumstances for such a construction to give any final standing wave. The
evaluation of $\psi_{\cal O}$ may be done [1] by using the intertwining
operator
(41) to calculate the Liouville Green function,
$$ G(z,w;\Delta t)=[C^{-1}e^{-\ft\im2\tilde p^2\Delta
t}C\delta(q-w)](z),\eqno(57)
$$ where $\Delta t=t-t_0$. In this way, one obtains the time-evolved
wavefunction
$$
\psi\subfor_{{\cal O}_p}(q,t)= 2^{\im
p}\int_{-\infty}^\infty{dk\over(2\pi)^2}\,
ke^{\pi k}K_{\im k}(e^q)
\Gamma\left(\im(p+k)\over2\right)\Gamma\left(\im(p-k)\over2\right) e^{-\ft\im2
k^2\Delta t}.\eqno(58)
$$

     The behavior of $\psi\subfor_{{\cal O}_p}(q,t)$ as $\Delta
t\rightarrow\infty$ may be evaluated by contour-integral methods [1]. Here, we
shall just summarize the results. The situation depends importantly on whether
$p$ is real or imaginary. The occurrence of imaginary momenta in noncritical
string theory is occasioned by the presence of background charges $Q$ as in
(19). In subcritical cases ($d<26$ for the ordinary string), the background
charges need to push the central charge of the compensating Liouville mode up
above its canonical value of 1, and in consequence, as one can see from (20),
the background charge is then real. In integrations in correlation functions
over the constant mode $\phi_0$ of the Liouville field, one then has at the
tree
level (where the Euler number of the worldsheet is 2) an extra factor
$e^{2Q\phi_0}$, as one can see from (19). This produces an extra ``background''
term of $-2\im Q$ in momentum-conservation delta functions, and makes the
consideration of imaginary momenta unavoidable. The normalizability
implications
of such imaginary momenta in the ``gravitational dressing'' of string states
are
not in our view yet fully established.

     For real $p$, there are different cases depending on whether $p\gtlt
0$:\par\narrow\noindent
 Real $p<0$: $\psi\subfor_{{\cal O}_p}\underbuildrel{\Delta
t\rightarrow\infty}\under\longrightarrow 0$ like $(\Delta t)^{-3/2}$;\nl Real
$p>0$: $\psi\subfor_{{\cal O}_p}\underbuildrel{\Delta
t\rightarrow\infty}\under\longrightarrow K_{\im
p}(e^q).$\par\nonarrower\noindent
Thus, for real $p<0$, the initial plane-wave wavefunction just dissipates in
expected for the generic case. For real $p>0$, however, the path-integral
implementation of the operator-state map (55) does work as desired and one is
left with an (improperly-normalized) standing wave proportional to a single
Liouville eigenstate (43).  The difference between the $p\gtlt 0$ cases may be
understood heuristically in terms of the need to set up a superposition of
incoming and outgoing plane waves in order to create a Liouville eigenstate.
For
$p>0$, this is possible owing to the entirely reflective nature of the
potential
$e^{2\phi}$. In this case, one has at $t_0\rightarrow -\infty$ an incoming wave
that subsequently reflects and produces an outgoing wave, with the
superposition
eventually settling down as $\Delta t\rightarrow\infty$ to a Liouville
eigenstate of the form (43). For $p<0$, on the other hand, the initial wave is
purely outgoing and so there is no way to generate the incoming wave that would
be needed to create a stationary state, so the wavefunction just dissipates as
generically expected, like
$(\Delta t)^{-3/2}$.

     For imaginary values $p=\im\beta$, there are again two cases depending on
whether $\beta\gtlt 0$:\par\narrow\noindent Imaginary $p=\im\beta$, $\beta<0$:
$\psi\subfor_{{\cal O}_p}\underbuildrel{\Delta
t\rightarrow\infty}\under\longrightarrow 0$ like $(\Delta t)^{-3/2}$;\nl
Imaginary $p=\im\beta$, $\beta>0$: $\psi\subfor_{{\cal
O}_p}\underbuildrel{\Delta
t\rightarrow\infty}\under\longrightarrow
\sum_{n=0}^{[\beta/2]}c_nK_{\beta-2n}(e^q)$,\nl where $[\beta/2]$ is the
integer
part of $\beta/2$.\par\nonarrower\noindent Thus, for $\beta<0$ one finds again
the generic case of a dissipating wavefunction. For $\beta>0$, however, one is
left in general with not one but a whole superposition of imaginary-momentum
Liouville eigenfunctions. The implications of this have not been fully worked
out, but the issue is important for the proper interpretation of Liouville
correlation functions, which have generally been considered using the
vertex-operator construction. The phenomenon of having only one sign of
momentum
give rise to a Liouville eigenstate is known as the Seiberg bound [19].
\bigskip
\leftline{\it Conclusions and Open Problems}

     The technique of solving Toda theory models via canonical transformations
implemented by intertwining operators highlights the similarities and
differences
between these integrable models and the free-field theories that are the basis
for conformal field theory. Although the intertwining-operator technique still
remains to be applied at the full field-theory level, indications on how that
may
be done can be obtained by comparison to B\"acklund transformation methods [20]
that have been successfully applied to Liouville field theory. The method of
Refs
[20] relies on an ans\"atz based upon the classical generating functional
$F(q,\tilde q)$ for the canonical transformation. In essence, that approach
expresses the interacting-theory wavefunction as an integral transform
involving
this generating functional,
$$
\psi(q)=\int d\tilde q\, e^{iF(q,\tilde q)}
\tilde\psi(\tilde q),\eqno(59)
$$ In promoting this transformation to the quantum case, one has to require
that
$e^{iF(q,\tilde q)}$ satisfy an analogue of our intertwining condition (37),
$$ H_{\rm L}(q,p)\,e^{\im F(q,\tilde q)}=\tilde H_{\rm L}(\tilde q,\tilde p)\,
e^{\im F(q,\tilde q)},\eqno(60)
$$ where the momenta are realized as derivatives in the Schr\"odinger
representation. In the case of Liouville theory, the classical generator
actually satisfies the condition (60) without further quantum corrections. This
could be related to the fact that our quantum transformation (39) turns out to
be one of the simple operator-ordering versions of the classical transformation
(35). Whether this luck will persist in the more general Toda cases remains to
be determined.

     From the BRST point of view, an open problem remains the role of the ghost
fields in the field-theoretic extension of the canonical transformations and in
the Weyl-group structure of these transformations. In the cases of free-field
Virasoro or $W_3$ gravities with minimal field content ({\it i.e.}\ just the
fields
$\phi$ or $\phi_{1,2}$), it is remarkable that when one includes the oscillator
states a Weyl-multiplet structures persists in the spectra, corresponding to
$Z_2$ or $S_3$ transformations of the center-of-mass mode momenta [21, 1]. But
these Weyl-group multiplets involve states of non-trivial ghost structure,
unlike
the situation at the minisuperspace level that we have considered here. This
suggests that in worldsheet gravity theories the Liouville- or Toda-theory
aspects cannot be completely disentangled from the gauge-theory aspects of the
problem. Another puzzle in the BRST context is the origin of hidden symmetries
such as the $SL(2,\R)$ Ka\v c-Moody symmetry of the correlation functions [6],
and how such symmetries might be related to the Weyl-group symmetries in the
canonical-transformation approach. Overall, it seems that unraveling the
mysteries of non-critical worldsheet gravity theories will require a more
profound synthesis of these different approaches.\np
\centerline{\bf REFERENCES}
\frenchspacing
\bigskip
\item{[1]}A. Anderson, B.E.W. Nilsson, C.N. Pope and K.S. Stelle, ``The
Multivalued Free-Field Maps of Liouville and Toda Gravities,'' {\tt
hep-th/9401007}, {\sl Nucl. Phys.} {\bf B}, in press.
\item{[2]}C. Teitelboim, {\sl Phys. Rev.} {\bf D25} (1982) 3159.
\item{[3]}J. Zinn-Justin, ``Renormalization of Gauge Theories,'' in {\it Proc.
Int. Summer Inst. for Theor. Physics (Bonn, 1974).}
\item{[4]}I.A. Batalin and G.A. Vilkovisky, {\sl Phys. Lett.} {\bf 102B} (1981)
27;\nl {\sl Phys. Rev.} {\bf D28 } (1983) 2567.
\item{[5]}C.M. Hull and P.K. Townsend, {\sl Nucl. Phys.} {\bf B301}1988 197.
\item{[6]}A.M. Polyakov,  {\sl Phys. Lett.} {\bf 101B} (1981) 207; {\sl Mod.
Phys. Lett.} {\bf A2} (1987) 893.
\item{[7]}F. David and E. Guitter, {\sl Euro Phys. Lett.} {\bf 3} (1987)
1169;\nl
F. David, {\sl Mod. Phys. Lett.} {\bf A3} (1988) 1651;\nl J. Distler and H.
Kawai, {\sl Nucl. Phys.} {\bf B321} (1989) 509.
\item{[8]}E. D'Hoker, {\sl Mod. Phys. Lett.} {\bf A}, (1991) 745.
\item{[9]}J. Benn, {\sl Phys. Lett.} {\bf B332} (1994) 329.
\item{[10]}A.B. Zamolodchikov, {\sl Teo. Mat. Fiz.} {\bf 65} (1985) 644.
\item{[11]}J. Thierry-Mieg, {\sl Phys. Lett.} {\bf 197B } (1987) 368.
\item{[12]}L.J. Romans, {\sl Nucl. Phys.} {\bf B352} (1991) 829.
\item{[13]}V.A. Fateev and A.B. Zamolodchikov, {\sl Nucl. Phys.} {\bf B280}
(1987) 644.
\item{[14]}S.R. Das, A. Dhar and S.K. Rama, {\sl Mod. Phys. Lett.} {\bf A6}
(1991) 3055; {\sl Int. J. Mod. Phys.} {\bf A7} (1992) 2295.
\item{[15]}J.-L. Gervais and A. Neveu, {\sl Nucl. Phys.} {\bf B199} (1982) 59;
{\bf B209} (1982) 125; {\bf B224} (1983) 329; {\bf B238} (1984) 125; {\bf B238}
(1984) 396; {\bf B257 [FS14]} (1985) 59; {\bf B264} (1986) 557; {\sl Commun.
Math. Phys.} {\bf 100} (1985) 15; {\sl Phys. Lett.} {\bf B151} (1985) 271;\nl
J.-L. Gervais, {\sl Commun. Math. Phys.} {\bf 130} (1990) 257; {\bf 138} (1991)
301; {\sl Phys. Lett.} {\bf B243} (1990) 85; {\sl Nucl. Phys.} {\bf B391}
(1993)
287;\nl J.-L. Gervais and J. Schnittger, {\sl Nucl. Phys.} {\bf B413} (1994)
433.
\item{[16]}H.J. Otto and G. Weigt, {\sl Phys. Lett.} {\bf B159} (1985) 341;
{\sl
Z. Phys.} {\bf C31} (1986) 219;\nl G. Weigt, {\sl Phys. Lett.} {\bf B277}
(1992)
79; ``Canonical quantization of the Liouville theory, quantum group structures
and correlation functions,'' in {\it Pathways to fundamental theories}, Proc.
Johns Hopkins Workshop on Current Problems in Particle Theory 16 (World
Scientific, 1993);\nl Y. Kazama and H. Nicolai, {\sl Int. J. Mod. Phys.} {\bf
A9} (1994) 667.
\item{[17]}A. Anderson and R. Camporesi, {\sl Commun. Math. Phys.} {\bf 130}
(1990) 61;\nl A. Anderson, {\sl Phys. Lett.} {\bf B319} (1993) 157; {\sl Phys.
Rev.} {\bf D47} (1993) 4458; {\sl Ann. Phys.} {\bf 232} (1994) 292.
\item{[18]}M.A. Olshanetsky and A.M. Perelomov, {\sl Phys. Rep.} {\bf 94}
(1983)
313.
\item{[19]}N. Seiberg, ``Notes on quantum Liouville theory and quantum
gravity,'' in {\it Common trends in mathematics and quantum field theory},
Proc.
of the 1990 Yukawa International Seminar, {\sl Prog. Theor. Phys.} Suppl. 102
(1991).
\item{[20]}T. Curtright, ``Quantum B\"acklund transformations and conformal
algebras,'' in {\it Differential Geometric Methods in Theoretical Physics}, eds
L.L. Chau and W. Nahm (Plenum Press, NY, 1990);\nl T. Curtright and G.I.
Ghandour, ``Using Functional Methods to Compute Quantum Effects in the
Liouville
Model,'' in {\it Quantum Field Theory,  Statistical Mechanics, Quantum Groups
and Topology,} eds T. Curtright, L. Mezincescu and R. Nepomechie (World
Scientific, Singapore, 1992).
\item{[21]}H. Lu, C.N. Pope, X.J. Wang and K.W. Xu, {\sl Class. Quantum Grav.}
{\bf 11} (1994) 967.
\bye